\documentclass[12pt]{article}
\usepackage{epsfig,graphicx,bm,amsmath}
\newcommand{\be}{\begin{equation}}
\newcommand{\bea}{\begin{eqnarray}}
\newcommand{\eea}{\end{eqnarray}}
\newcommand{\ba}{\begin{array}}
\newcommand{\ea}{\end{array}}
\newcommand{\ee}{\end{equation}}
\expandafter\ifx\csname mathbbm\endcsname\relax

\else

\fi \textheight 22cm \textwidth 15cm \topmargin 1mm \oddsidemargin
5mm \evensidemargin 5mm

\begin{document}

\begin{titlepage}
\hfill

\vspace*{13mm}
\begin{center}
{\Large {\bf Non-Supersymmetric Unattractors in Born-Infeld Black Holes}\\
} \vspace*{18mm} \vspace*{1mm} {Sangheon Yun}
\footnote{sanhan1@phya.snu.ac.kr}\\
\vspace*{1cm} { \vspace{3mm} School of Physics and Astronomy, Seoul National University\\
Seoul 151-747, KOREA \\
\vspace*{6mm} }
\end{center}
\begin{abstract}
We investigate unattractor behavior in non-extremal black holes in
Einstein-Born-Infeld-Dilaton theory of gravity in four-dimensional
asymptotically flat spacetime. We obtain solutions which are
non-singular near the horizon and dependent on the value of the
dilaton field at the infinity, using perturbation method. It is
shown that the value of the scalar field at the horizon is
determined by its asymptotic value and the charges carried by the
black hole. And we also find it is not true in general that the
dilaton value at the horizon is a monotonically increasing
function of the first coefficient of its series expansion in
non-extremal Born-Infeld black holes.
\end{abstract}
\end{titlepage}

\section{Introduction}

In extremal black holes, the attractor mechanism states that the near horizon geometry, the field configuration of massless scalars and the black hole entropy turn out to
be completely independent of the asymptotic values of radially
varying scalar fields of the theory and dependent only on certain
conserved quantities like mass, charges and angular momentum \cite{0507096}-\cite{9603090}. It was
discovered first in N=2 extremal black holes
\cite{9508072}-\cite{9603090}, but the concept of attractor
mechanism was found to work in a much broader context and generalized beyond the original
key ingredient, supersymmetry. This
mechanism can be used to study the properties of extremal black
holes in supersymmetric theories, or in non-supersymmetric
ones. Examples of it also include black holes with higher derivative corrections, extremal black holes in higher dimensions and rotating black holes. Recently, Sen proposed, so called, the entropy function formalism
\cite{0506177} which is proved to be very
useful in calculating the entropy of extremal black holes in a
general theory of gravity, with any set of higher derivative terms
and in higher dimensions \cite{0007195}-\cite{Sahoo:2006pm}. This
formalism is based on the facts that the near horizon
geometries of the black holes are maximally symmetric and enough to give the entropy by the
Wald's entropy formula, and that attractor equations are essentially
some linear combinations of the equations of motion of all the
fields of the theory.

\vskip 0.3cm

Understanding the structure of the higher derivative terms is
crucial because they hold a lot of information about the unitarity
and renormalizability properties of the theory in question. With
attractor mechanism and entropy function
formalism, a lot of interesting aspects of Lovelock terms,
Chern-Simons terms, Born-Infeld terms etc., can be studied
\cite{0611240}, \cite{0511306}-\cite{0604028}. They are also important from the
point of view of the need to introduce a small amount of
non-extremality, in certain situations involving higher derivative
terms~\cite{0611143}. So far, compared with extremal black holes,
non-extremal ones are relatively less investigated. Recently, more
attention has been paid to unattractor and entropy function for
non-extremal black
holes~\cite{0507096},\cite{Morales:2006gm},\cite{0611140}-\cite{0701158}. Since the Einstein field equations are a set of second order coupled nonlinear partial differential equations which are free of time derivatives in this letter, we expect that initial conditions including the first coefficient of the scalar series expansion determine all the coefficients of the field series expansion. So it is natural that non-extremal black holes show unattractor behavior. In this paper, we pay more attention to the 'degree of unattractor' and find that our
naive intuition for the first coefficient of the scalar field series
expansion to increase monotonically with its value at the
horizon does not hold generically when Born-Infeld term is
considered.

\vskip 0.3cm

In this note, we review attractor mechanism and study unattractor
in Einstein-Born-Infeld theory of gravity coupled to a massless
neutral moduli field. Born-Infeld terms are known to arise in the
low energy limit where gauge fields are coupled
to open bosonic strings or superstrings. In fact, the low-energy effective theories of the $D$-brane worldvolumes are governed by Born-Infeld actions. The importance of Born-Infeld terms on the
black holes and connection with elementary string states was
stressed in~\cite{9506035}. It was argued that virtual black holes
going around closed loops can give rise to Born-Infeld type
corrections to extremal black hole configurations.
Einstein-Born-Infeld black holes in presence of string generated
low energy fields have been studied
in~\cite{Wiltshire:1988uq}-\cite{0101083}.

\vskip 0.3cm

The rest of this paper is organized as follows. In section 2, we
start with some relevant features of attractor mechanism needed
for our purposes in the case of  Einstein-Maxwell theory coupled
to massless neutral real scalar fields. Section 3 is devoted to
reviewing attractor mechanism in Einstein-Born-Infeld theory
coupled to a massless neutral real scalar field, with different value of $\lambda$ from the previous paper \cite{0611240}. But we make use of a
different basis. Then we investigate unattractor solutions in
Einstein-Maxwell theory in section 4, and generalize them to
Einstein-Born-Infeld theory in section 5, where we will investigate the degree of unattractor via the first coefficient of the scalar field series expansion. Our conclusions are summarized in section 6.

\section{Non-Supersymmetric Attractors: General Features }

Let us start with a few relevant aspects of non-supersymmetric
attractors needed for our purposes. We consider the class of
following gravity theories coupled to $U(1)$ gauge fields and
scalar fields as in \cite{0507096}\footnote{In the convention of
\cite{0507096}, the factor of $\frac{1}{\sqrt{-g}}$ is involved in
the definition of $\epsilon^{\mu \nu \rho \sigma}$.}: \be
\label{EHaction}
  S=\frac{1}{\kappa^{2}}\int d^{4}x\sqrt{-g}\left(R-2 \partial_\mu \phi_i \partial^\mu \phi^i-
  f_{ab}(\phi_i)F^a_{\mu \nu} F^{b \ \mu \nu} - \frac{1}{2\sqrt{-g}}{\tilde f}_{ab}(\phi_i) F^a_{\mu \nu}
  F^b_{\rho \sigma} \epsilon^{\mu \nu \rho \sigma} \right)
 \ee
where $F^a_{\mu\nu}, \; a=0,...N$ are gauge field strengths and
$\phi_i, \; i=1,...n$ are scalar fields. Since classical
non-abelian fields have never been observed, we will not take them
into account in this paper. The scalar-dependent couplings of
gauge fields are motivated from analogy with the supersymmetric
theories. Any additional potential term for the scalar fields will
lead to a breakdown of attractor mechanism in asymptotically flat
spacetimes which we consider here. Rest of the notations are as in
\cite{0507096}.

A static spherically symmetric ansatz is: \bea \label{metric}
ds^2&=&-\alpha(r)^2dt^2+\alpha(r)^{-2}dr^{2}+\beta(r)^2d\Omega^2
\eea On the other hand, the Bianchi identity and equations of
motion of gauge fields can be solved by taking the gauge field
strengths to be of the form: \be \label{gaugefield}
F^a=f^{ab}(\phi_i)(Q_{eb}-\tilde{f}_{bc}Q^c_m)\frac{1}{\beta^2}dt\wedge
dr + Q^a_m \sin\theta d\theta \wedge d\varphi, \ee where $Q^a_m$
and $Q_{ea}$ are constants that determine the magnetic and
electric charges carried by the gauge fields $F^a$, and $f^{ab}$
is inverse of $f_{ab}$. It is worth noting that the equations of
motion (except the Hamiltonian constraint which must be imposed
additionally\footnote{Otherwise, the Hamiltonian constraint can
also be derived from the one-dimensional action where we replace
the metric $g_{rr}$ component $\frac{1}{\alpha(r)^2}$ by
$\frac{1}{\eta(r)^2}$, take the variation with
respect to $\eta(r)$ and then set $\eta(r)$ equal to $\alpha(r)$.}) can be derived from the following
one-dimensional action:\be S = \frac{1}{\kappa^2} \int dr \left[
2-(\alpha^2\beta^2)^{''}-2\alpha^2\beta\beta^{''}-2\alpha^2\beta^2\phi^{'2}-\frac{2V_{eff}(\phi)}{\beta^2}\right]\ee
with the effective potential given by \be \label{effective}
V_{eff}(\phi_i) =
f^{ab}(Q_{ea}-\tilde{f}_{ac}Q^c_{m})(Q_{eb}-\tilde{f}_{bd}Q^d_{m})
+ f_{ab}Q^a_{m}Q^b_m \ee

Now two sufficient conditions for the moduli fields to have the
attractor behavior can be stated as follows \cite{0507096}. First,
for given charges, $V_{eff}$ as a function of the moduli, must
have a critical point $\phi_{i0}$. Then we have \be \partial_i
V_{eff}(\phi_{i0})=0\ee Second, the matrix of second derivatives
of the potential at the critical point, \be
M_{ij}=\partial_i\partial_jV_{eff}(\phi_{i0})\ee,
should have positive eigenvalues. Roughly we can write \be
M_{ij}>0\ee This condition guarantees the stability of the
solutions. Once the two conditions are met, the attractor
mechanism typically works \cite{0507096}. As discussed in \cite{0511117}, it is possible that some of the eigenvalues of $M_{ij}$ vanish. In that case the leading correction to $V_{eff}$ along a zero mode directions should be positive for the attractor behavior to exist.

\vskip 0.3cm

Let us concentrate on the scalar field equations of motion: \be
\partial_r(2\alpha^2\beta^2\partial_r\phi_i) = \frac{1}{\beta^2} \, \partial_i V_{eff} \label{eqphi} \\
\ee Non-supersymmetric attractor equations can be derived from
$\partial_i V_{eff}(\phi_{i0})=0$, which also determines the
attractor values of scalar fields in terms of the fixed charges of
the extremal black hole.

\vskip 0.3cm

As discussed in \cite{Chandrasekhar:2006kx}, the above analysis
can be generalized to include a certain set of higher derivative
terms coming from the gravity side. In other words, it was argued
in \cite{Chandrasekhar:2006kx} that, in the presence of general
$R^2$ terms in the action, the effective potential gets modified
by additional terms, and was in fact called as $W_{eff}$. The
scalar field equation of motion remains as in (\ref{eff}), with
$V_{eff}$ replaced by $W_{eff}$.

\vskip 0.3cm

Here, it should be mentioned that $W_{eff}$ will in general depend
on $r$. However, near the horizon all the quantities are
independent of $r$. In this special situation the $r$ dependence
in $W_{eff}$ drops out. As a result, the horizon radius computed
from $W_{eff}$ will also be a constant, but modified by higher
derivative terms. For instance, let us note down the general form
of the scalar field equation near the horizon, in the presence of
Gauss-Bonnet terms: \be \left(2 \alpha^2 \beta^2\phi'\right)' =
\frac{1}{\beta^2}\frac{d W_{eff}}{d\phi} \ee where \be
W_{eff}(\phi) = V_{eff}(\phi) + 4\,G(\phi) \, \ee and there
is no $r$ dependence. The additional term $4\,G(\phi)$ also
modify the entropy of the black hole via Wald's entropy formula.
This is parallel to the analysis in Sen's entropy function
formalism, where the addition of  Gauss-Bonnet term gives rise to
a finite horizon area and entropy of small black holes.

\vskip 0.3cm

Black hole solutions in Einstein-Born-Infeld theories have been
studied quite a lot in literature. It is known that one can have
particle-like and BIon solutions in these theories. However,
finding explicit black holes solutions in the presence of scalar
couplings in the Einstein-Born-Infeld action is non-trivial. In
four dimensions, when looking for asymptotically flat solutions in
these theories, it is reasonable to assume that the near horizon
geometry of these black holes preserve the symmetries of $AdS_2
\times S^2$. In order to understand the effect of higher order
Born-Infeld corrections to the entropy of extremal black holes, an
entropy function analysis of small black holes in heterotic string
theory was presented in \cite{0604028}. However, it
is important to check if the attractor mechanism works when
considering the full black hole solution. Later we will discuss
this point in more detail. As in
\cite{0507096,Chandrasekhar:2006kx}, in this work, we carry out a
perturbative analysis to show that the moduli fields take fixed
values as they reach the horizon and that a double horizon
Einstein-Born-Infeld black hole continues to exist. We show that
the attractor mechanism works in the case of Born-Infeld black
holes. In effect, we show that once one obtains critical values of
the effective potential and ensure that $\partial_i
\partial_j V_{eff}(\phi_0) > 0$, the perturbative analysis signifies
that the attractor points remain stable.

\section{\bf Non-Supersymmetric Attractors in Einstein-\\Born-Infeld Theories}

Non-supersymmetric attractor mechanism in Einstein-Born-Infeld
theories can be studied using the entropy function formalism.
However, to see the moduli indeed get attracted to fixed points
near the horizon, one has to use the formalism for
non-supersymmetric attractor mechanism reviewed in the previous
section, which make explicit use of the general solution and
equations of motion \cite{0507096}. In this section, we follow the
analysis outlined in the previous section and the recent
paper~\cite{0611240}. Using a perturbative approach to study the
corrections to the scalar fields and taking the backreaction
corrections into the metric, it is possible to show that the
scalar fields are indeed drawn to their fixed values at the
horizon. Here, the requirements are the existence of a {\it double
degenerate horizon solution}. The existence of a
non-supersymmetric attractor mechanism for higher derivative
gravity has been recently studied in~\cite{Chandrasekhar:2006kx}.
\vskip 0.2cm

Thus, it should be interesting to use the equations of motion and
study the attractor mechanism in the case of Einstein-Born-Infeld
black holes coupled to moduli fields. For the purpose of studying
non-supersymmetric attractor mechanism, it is instructive to start
from the following action: \be \label{last}
S=\frac{1}{16\,\pi}\int d^{4}x\sqrt{-g}\left(R_g - 2\,{\tilde
g}_{ij}\partial\phi^i\partial\phi^j + {\cal L}_{BI}^{(a)} \right)
\ee
where\\
\be {\cal L}_{BI} = 4b f_{(a)}(\phi_i)\left\{1  -  \left[1 +
\frac{f^{2(a)}(\phi)}{2b}F^2  - \frac{f^{4(a)}(\phi)}
{16b^2}(F\star F)^2  \right]^\frac{1}{2} \right\}. \ee where we restricted to four spacetime dimensions and used the static (or Monge) gauge, $i$
runs over the number of scalars and $a$ is the number of gauge
fields, with $F^a_{\mu \nu}$ denoting the field strength, $\star F$ dual to the Maxwell tensor $F$ and
$f_{(a)}(\phi_i)$ determining the couplings. It is important not
to have a potential for the scalar fields, so as to allow for a
moduli space to vary. Here, ${\tilde g}_{ij}$ stand for the metric
of the moduli space. In the string theory, the
parameter $b$ is related to the inverse string tension $\alpha '$ as $b = (2 \pi \alpha ')^{-2}$. Note that the
action reduces to the Maxwell system in the $b\rightarrow\infty$ limit. We continue to retain the Born-Infeld parameter $b$. In the absence of any moduli fields, Einstein-Born-Infeld black holes have been constructed in
\cite{Wiltshire:1988uq}. In what follows, we will be interested in
asymptotically flat spacetime solutions, although the
generalization to include a cosmological constant should also be
possible. In fact, it might be interesting to include a
cosmological constant~\cite{Dey:2004yt} in view of the results in
\cite{0604028}.

\subsection{Single Scalar and Gauge Field Case}

 \vskip 0.2cm Let us take
$f_{(a)}(\phi_i)=e^{2\gamma_a \phi_i}$ where $\gamma_a$ are
parameters characterizing the coupling strength of dilaton field.
It is one for string theory. We keep this parameter, for an
arbitrary value of this parameter is possible in a general theory
of gravity in four dimensions.

\vskip 0.2cm

Now, one makes an ansatz for a static spherically symmetric metric
which must satisfy the field equations following from the
Einstein-Born-Infeld action in eqn. (\ref{last}). For simplicity,
we restrict ourselves to the single scalar and gauge field
case\footnote{The generalization to multi-scalar fields is
straightforward}. One can generalize the result when we have
several scalars and gauge fields. It should be mentioned that,
although we are working with a system of gauge fields coupled to
scalar fields, to lowest order, we are looking for the solution of
the equations of motion only for constant values of moduli. The
Birkhoff's theorem holds in this case and we may assume the
solution to be static and spherically symmetric, to be of the
form: \bea ds^{2} &=&
-\alpha(r)^{2}\,dt^{2}\,+\;\frac{dr^{2}}{\alpha(r)^{2}}\;+\;
\beta(r)^{2}\;d\Omega^{2}_2  \crcr F &=& F_{tr} \; dt \wedge dr +
F_{\theta\phi}\;d\theta \wedge d\phi. \eea The induction tensor
$G_{\mu\nu}$ is defined by \be G^{\mu\nu} = -\frac{1}{2}
\frac{\partial L}{\partial F_{\mu\nu}} \ee The Maxwell equations
and Bianchi identity are \be dG=0 \hspace{2cm} dF=0 \ee These give
us the following solution \be F_{tr}= \frac{Q_e
e^{2\gamma\phi}}{\beta^2
\sqrt{1+\frac{Q^{2}_e+Q^{2}_me^{-4\gamma\phi}}{b\beta^4}}},
\hspace{10mm} F_{\theta\phi}= Q_m\sin\theta \ee Although, for
simplicity, we consider the case of a single scalar field, the
generalization to many scalar fields is straightforward. The
equations of motion and Hamiltonian constraint are driven from the
action $S$ with the above solution for gauge fields  and metric
ansatz, as follows \bea \label{eqn1} -1 + \left(\frac{\alpha^2\beta^{2'}}{2}\right)' +
\frac{1}{\beta^2}V_{eff} &=& 0 \\ \cr\cr \label{ab}
\left(\frac{\alpha^{2'}\beta^2}{2}\right)'-2 b \beta^2
e^{2\gamma\phi}\left(1-\frac{1}{\sqrt{1+\frac{Q_e^2+Q_m^2e^{-4\gamma\phi}}{b\beta^4}}}
\right) &=& 0\\ \cr\cr \label{dil}
\partial_{r}(2\alpha^{2}\,\beta^{2}\partial_{r}\phi)
- \frac{\partial_{\phi}V_{eff}}{\,{\beta^2}} &=& 0 \\ \cr
\label{eqn4} (\partial_r \phi)^2  + \frac{\beta''}{\beta}&=& 0
\eea where $V_{eff}$ plays a role of an `effective potential' for
the scalar fields. A difference with \cite{0507096} is that in
this case, $V_{eff}$ is a function of $r$, as seen below: \be
\label{eff} V_{eff} = 2b\,\beta^4 e^{2\gamma\phi}\left(
{\sqrt{1+\frac{Q_e^2+Q_m^2e^{-4\gamma\phi}}{b\beta^4}}} -1 \right)
\ee However, as discussed in~\cite{0601016}, it is possible to
treat $r$ as just a parameter near the horizon. Extremizing the
effective potential gives the fixed values taken by the moduli at
the horizon.

\subsection{Perturbative Analysis}

\vskip 0.2cm

It is well known that these equations admit $AdS_2 \times S^2$ as
a solution in the case of constant moduli. However, we wish to
address the attractor behavior considering double horizon black
hole solutions, which are asymptotically flat. Thus, we start with
an extremal black hole solution in this theory, obtained by
setting the scalar fields at their critical values of the
effective potential. Then, as one varies the values of scalar
fields at asymptotic infinity, we show that the double horizon
nature of black hole remains. Further, the critical values of the
scalar fields remain stable, as the asymptotic values of these
moduli fields are somewhat different from attractor values.

\vskip 0.2cm

In view of the fact that the four equations governing
$(\alpha(r),\beta(r),\phi(r))$ are a set of four highly
complicated coupled differential equations, we follow the
Frobenius method to solve these equations, but exploit different
basis from~\cite{Chandrasekhar:2006kx} and~\cite{0611240}. We call
these four sets of equations of motion $(EqA,EqB,Eq\Phi,EqC)$. As
a variable of expansion we define $x \equiv (\frac{r}{r_H}-1)$.
Requiring that the solution: (a) be extremal: meaning that we have
double horizon\footnote{Later we will see that there is no
attractor mechanism in a single-zero horizon case. Thus a
double-zero horizon is important for the presence of the attractor
mechanism.}, therefore $\alpha^2(r)=(r-r_H)^2\tilde\alpha^2(r)$,
with $\tilde\alpha^2(r)$ being analytic at the horizon, $r=r_H$,
(b) be asymptotically flat: meaning that geometry and moduli tend
to flat geometry at asymptotic infinity and (c) be regular at the
horizon\footnote{Just like the case which has been studied in
\cite{0507096} there is a solution where the scalar blows up at
the horizon. In the supersymmetric case, the well behaved solution
is automatically chosen.}, the most general Frobenius expansions
of $\alpha(r)$,$\;\beta(r)$ and $\phi(r)$ take the form:
\bea\label{aexpan} \alpha^2(r)&=&\; \alpha_H^2 x^2
\sum_{m,n=0}^{\infty}a_{m,n} x^{m \lambda +n}
,  \\
\label{bexpan}
\beta(r)&=&\;  r_H \sum_{m,n=0}^{\infty} b_{m,n}x^{m \lambda +n} ,  \\
\label{phiexpan} \phi(r)&=&\; \sum_{m,n=0}^{\infty}\phi_{m,n} x^{m
\lambda +n}, \eea with $\lambda \geq 1$,\; $a_{0,0}=1$,\; $b_{0,0}=1$ and
$\phi_{0,0}=\phi_{0}$. In
\cite{0507096} and \cite{0611240}, $\lambda$ was assumed to be
very tiny positive number, $\lambda \ll 1$. In the above expansion, however, we assumed $\lambda \geq 1$
for $\partial_{r}\phi$ not to diverge near the horizon.\vskip 0.2cm

We mention that although, in comparison to (\ref{eff}), $V(\phi)$
of (\ref{eff}) is of pure magnetic (electric) type, the case given
in (\ref{eff}) does not have a minimum for any finite value of
$\phi$ in pure electric or magnetic case. To have a minimum in
single charge case we need at least two gauge fields. Here we
consider dyonic case where both electric and magnetic charges are
non-zero, and assume that $\lambda$ is slightly greater than 1 from now on.

\begin{flushleft}
\underline{{\bf Zeroth order results}}
\end{flushleft}

At zeroth order perturbation we start with a double horizon black
hole solution as follows: \bea\label{v0} \phi(r) = \phi_0,
\;\;\;\;\;\;\ \beta(r) = r_H, \;\;\;\;\;\;\;\; \alpha(r) = \alpha_H
(\frac{r}{r_H}-1) \eea where for given electric and magnetic
charges, $\phi_0$, $\alpha_H$ and $r_H$, can be found from the
following equations in terms of these charges\footnote{Another useful relations are as follows:\bea
2br_H^2e^{2\gamma\phi_0} = \frac{\alpha_H^2}{1-\alpha_H^2}, \;\;\;\;\;\;\;\; 1+\frac{Q_e^2}{br_H^4} = \frac{1}{\alpha_H^2}, \;\;\;\;\;\;\;\;
 4 b Q_m^2 = \frac{1}{1-\alpha_H^2} \nonumber
\eea}
\bea
e^{4\gamma\phi_0} &=& \frac{Q_m^2}{Q_e^2}\alpha_H^2 \\
\cr r_H^4 &=& 4 Q_e^2 Q_m^2\alpha_H^2 \\ \cr \label{v1}
 \alpha_H^2 &=& 1- \frac{1}{4 b Q_m^2}
\eea As seen above, $\alpha_{H}$ is less than 1.

We should mention that from the above equations we find a lower
bound for magnetic charge value $4b Q_m^2 > 1$\footnote{Note the proposal in \cite{0101083} about the non-existence of the extremal limit for electrically charged black holes with Born-Infeld term.
}. This bound relaxes in the limit
$b\rightarrow \infty$ where Born-Infeld theory reduces to Maxwell
theory. In this limit $\phi_0$ and $r_H$ approach values that one
can find in Einstein-Maxwell-Dilaton theory~\cite{0507096}. In
this case, a Reissner-Nordstrom  black hole with constant scalars,
is an exact solution of the equations of motions. \vskip 2mm

Notice that the equations (\ref{v0}-\ref{v1}), together, determine
both the attractor value of the moduli field and the horizon
radius in terms of charges and the parameters of the action. In
fact both the above results are meaningful. Due to (\ref{v0}) the
Bekenstein-Hawking entropy of the solution is given by the value
of the $V_{eff}(\phi_0)$, up to a numerical prefactor.

\vskip 0.2cm

This in fact fixes $\phi_0 $ at its extremum point. From
(\ref{phiexpan}), $\phi_0 = \phi(r_H)$ and so the value of the
moduli field is fixed at the horizon, regardless of any other
information. Thus to complete the proof of the attractor behavior,
we should be able to show that the four sets of equations of
motion, denoting a coupled system of differential equations, admit
the expansions (\ref{aexpan}), (\ref{bexpan}) and
(\ref{phiexpan}). Furthermore, one should see that there are
solutions to all orders in the $x$-expansion with arbitrary
asymptotic values at infinity, while the value at the horizon is
fixed to be $\phi_0$. The existence of a complete set of solutions
with desired boundary conditions (considering the fact that we
have coupled non-linear differential equations) by itself is not
trivial. Moreover, it is easy to show that, in our theory, there
is no asymptotically flat solution with everywhere constant
moduli.

\begin{flushleft}
\underline{{\bf First order results}}
\end{flushleft}

To start with first order perturbation theory, we write \be \delta
\phi \equiv \phi-\phi_0, \ee where we keep $\delta \phi $ as a
small parameter in perturbation. From the scalar equation of
motion we find \be \label{phi} \delta \phi = \phi_{1,0}
(\frac{r}{r_H}-1)^{\lambda}+ \frac{\gamma
(1-\alpha_H^2)}{\gamma^2-1}(\frac{r}{r_H}-1)\ee where $\phi_{1,0}$
is an undetermined constant and $
\lambda=\frac{1}{2}(-1+\sqrt{1+8\gamma^2})$. We see from eq.
(\ref{phi}) that $\delta \phi$ vanishes at the horizon and the
value of the scalar is fixed at $\phi_0$ regardless of its
asymptotic value. This shows that attractor mechanism works to the
first order in perturbation theory. It is noteworthy that the
first order solution for the scalar field has a term with power 1
as well as general $\lambda$-power one. The integer-power term at
the first order of the scalar field expansion appears from the
fact that the first order term in $\beta$-series expansion has
power 1. It really vanishes in the limit $\alpha_{H}\rightarrow1$
where Born-Infeld theory reduces to Maxwell one, as expected.

\vskip 0.2cm

In comparison to the Einstein-Maxwell theory where an
Reissner-Nordstrom black hole case was considered
in~\cite{0507096}, here we have corrections to metric components
at the first order in perturbation theory. At this order,
$\alpha^2(r)$ and $\beta(r)$ receive corrections as follows: \bea
\alpha_1(r)&=&\alpha_H^2 a_{1,0} (\frac{r}{r_H}-1)^{\lambda +2}+
\alpha_H^2 a_{0,1}(\frac{r}{r_H}-1)^3 \\
\cr \beta_{1}(r)&=&r_H(\frac{r}{r_H}-1)\eea where \bea a_{1,0}
&=&\frac{4\gamma (1-\alpha_H^2)}{(\lambda +1)(\lambda +2)}
\phi_{1,0} \\ \cr a_{0,1}&=&\frac{2(1-\alpha_H^2)^2}{3(\gamma^2
-1)}-2\alpha_H^2 \eea This correction
vanishes at the horizon faster than $(\frac{r}{r_H}-1)^2$. Thus to
this order, the solution continues to be a double horizon black
hole with vanishing surface gravity.

\vskip 0.3cm

\begin{flushleft}
\underline{{\bf Second order results}}
\end{flushleft}

At second order in perturbation theory the non-constant value of
scalar field we found at first order, plays the role of a source.
This results in corrections to the metric components.  We should
also consider boundary conditions as follows. Since we are
interested in extremal black hole solutions with vanishing surface
gravity, we should have a horizon where $\beta(r)$ is finite and
$\alpha^2(r)$ has a "double horizon". In other words,
$\alpha(r)=(r-r_H) \tilde\alpha(r)$ where $\tilde\alpha(r)$ is
finite and non-zero at horizon. It is useful to note that, by an
appropriate  gauge choice, we can always take the horizon to be at
$r=r_H$.  Plugging (\ref{aexpan}-\ref{phiexpan}) into equations
(\ref{eqn1}-\ref{eqn4}), the solutions for $\alpha$,$\beta$ and
$\phi$ corresponding to the above boundary conditions are \bea
\alpha_2(r) &=& \alpha_H^2 a_{2,0} (\frac{r}{r_H}-1)^{2\lambda +2}
+\alpha_H^2 a_{1,1}(\frac{r}{r_H}-1)^{\lambda +3}+\alpha_H^2
a_{0,2}(\frac{r}{r_H}-1)^4
\\ \cr \beta_2(r) &=& b_{2,0} r_H (\frac{r}{r_H}-1)^{2\lambda}
+b_{1,1} r_H (\frac{r}{r_H}-1)^{\lambda +1}+b_{0,2} r_H
(\frac{r}{r_H}-1)^{2}
\\ \cr \phi_2(r) &=& \phi_{2,0}(\frac{r}{r_H}-1)^{2\lambda}+\phi_{1,1}(\frac{r}{r_H}-1)^{\lambda
+1}+\phi_{0,2}(\frac{r}{r_H}-1)^{2}\eea where some of the
coefficients are \bea a_{2,0}&=& \frac{2\left(
1-\alpha_H^2-\alpha_H^4 -\lambda(2\lambda +3)\right)}{(\lambda
+1)(2\lambda +1)}b_{2,0}+\frac{2\gamma (1-\alpha_H^2)}{(\lambda
+1)(2\lambda +1)} \phi_{2,0}
\\ \cr b_{2,0}&=& -\frac{\lambda}{2(2\lambda -1)}\phi_{1,0}^2
\\ \cr b_{1,1}&=& -\frac{\lambda(1-\alpha_H^2)}{\gamma(\gamma^2 -1)}\phi_{1,0}
\\ \cr b_{0,2}&=& -\frac{1}{2}\left( \frac{\gamma (1-\alpha_H^2)}{\gamma ^2 -1}\right)^2
\\ \cr \phi_{2,0} &=&[ \frac{2\lambda(2\lambda+1)}{\gamma(\lambda+1)(\lambda+2)}
-\frac{\lambda}{2\gamma(2\lambda-1)}] (1-\alpha_H^2)\phi_{1,0}^2
\eea

These solutions vanish at the horizon. With vanishing of
$\beta_1(r)$, horizon area does not change to the second order in
perturbation theory and is therefore independent of the asymptotic
value of dilaton. $\alpha_2(r)$ also vanishes at the horizon
faster than $\alpha_1(r)$ thus the second order solution continues
to be a double horizon black hole with vanishing surface gravity.

\vskip 0.2cm The scalar also gets a correction to the second order
in perturbation. This can be calculated in a way similar to the
above analysis. We discuss this correction along with higher order
corrections.

\begin{flushleft}
\underline{{\bf Higher order results}}
\end{flushleft}

We solve the system of equations $(EqA,EqB,Eq\Phi,EqC)$ order by
order in the $x$-expansion. To first order, we find that one
variable, say $\phi_{1,0}$, cannot be fixed by the equations. Let
us denote the value of $\phi_{1,0}$ as $K$. We thus find $a_{1,0}$
and $b_{1,0}$ as functions of $K$. One can check that at any order
$n \geq 2$, one can substitute the resulting values of
$(a_{m,l},b_{m,l},c_{m,l})$, for all $m+l \leq n$ from the
previous orders.  Then $(EqB,Eq\Phi,EqC)$ of the current order
together with $EqA$ of order $(n-1)$, consistently give
\bea\label{resultn}
b_{n,l}=b_{n,l}(K)\;\;\;;\;\;\;a_{n,l}=a_{n,l}(K)\;\;\;;\;\;\;\phi_{n,l}=\phi_{n,l}(K)\;\;\;.
\eea as polynomials of order $n$ in terms of $K$. \vskip 0.2cm $K$
remains a free parameter to all orders in the $x$-expansion. From
(\ref{aexpan}), (\ref{bexpan}) and (\ref{phiexpan}), the
asymptotic values of $(\alpha(r),\beta(r),\phi(r))$ are given by a
sum of all the coefficients in the x-expansion of the
corresponding function. After changing bases from
$(\frac{r}{r_H}-1)$ to $(1-\frac{r}{r_H})$, as a consequence of
(\ref{resultn}), one notices that
$(\alpha_\infty,\beta_\infty,\phi_\infty)$ are free to take
different values, given different choices for $K$. The convergence
of the series is not addressed in detail, but it would be the case
for small enough values for $|K|$. \vskip 0.2cm The arbitrary
value of $\phi$ at infinity is $\phi =\phi_\infty$, while its
value at the horizon is fixed to be $\phi_0$. This signifies the
presence of attractor mechanism in this theory.

\section{Non-Supersymmetric Unattractors in Einstein-Maxwell Theories}

For simplicity, we take $f(\phi)=e^{2\gamma \phi}$ where $\gamma$
is a parameter characterizing the coupling strength of the dilaton
field. We keep this parameter, as its arbitrary value is possible
in a general gravitational theory in four dimensions.

Again we assume the solution to be static and spherically
symmetric, to be of the form: \bea ds^{2} &=&
-\alpha(r)^{2}\,dt^{2}\,+\;\frac{dr^{2}}{\alpha(r)^{2}}\;+\;
\beta(r)^{2}\;d\Omega^{2}_2  \crcr F &=& F_{tr} \; dt \wedge dr +
F_{\theta\phi}\;d\theta \wedge d\phi. \eea The Maxwell equations
and Bianchi identity give us the following solution \be F_{tr}=
\frac{Q_e e^{2\gamma\phi}}{\beta^2}, \hspace{10mm} F_{\theta\phi}=
Q_m\sin\theta \ee The equations of motion and Hamiltonian
constraint are driven from the action $S$ with the above solution
for gauge fields  and metric ansatz, as follows \bea \label{eqn1}
-1 + \left(\frac{\alpha^2\beta^{2'}}{2}\right)'+ \frac{1}{\beta^2}V_{eff} &=& 0 \\
\cr\cr \label{ab} \left( \alpha^2\beta^2 \right)'' &=& 2\\
\cr\cr \label{dil}
\partial_{r}(2\alpha^{2}\,\beta^{2}\partial_{r}\phi)
- \frac{\partial_{\phi}V_{eff}}{\,{\beta^2}} &=& 0 \\ \cr
\label{eqn4} (\partial_r \phi)^2  + \frac{\beta''}{\beta}&=& 0
\eea where $V_{eff}$ is an `effective potential' for the scalar
field. As in \cite{0507096}, $V_{eff}$ in this case, is a function
of $\phi(r)$, as seen below: \be \label{eff} V_{eff} = Q_{e}^2
e^{2\gamma\phi}+Q_{m}^2 e^{-2\gamma\phi} \ee In this section we
are considering non-extremal black holes. From the equation (44),
we have $\alpha^2\beta^{2}=(r-r_{+})(r-r_{-})$ where
\;$r_{+}=r_{-}$\; for extremal black holes and \;$r_{+}>r_{-}$\;
for non-extremal cases which are of interest to us at the moment.

In perturbation, at the first order, the equation for the scalar
field $\phi=\phi_0 + \delta\phi$ takes the form: \be \partial_r
(\alpha^2 \beta^2 \partial_r
\delta\phi)=\frac{1}{2\beta^2}V_{eff}''(\phi_0)\delta\phi\ee In
the vicinity of the horizon $r=r_+$, this is given by \be
\partial_x[x(x+1-\frac{r_-}{r_+})\partial_x \delta\phi]=\frac{4\gamma^2 |Q_e Q_m|}{r_+ ^2}\delta\phi\ee
Using change of variables $x=-(1-\frac{r_-}{r_+})y$, we find the
solution is proportional to a hypergeometric function:\be
\delta\phi=C_0 + C_1y + C_2 y^2 + \cdots\ee where the ellipses
indicate the higher order terms in the expansion of $\delta\phi$
around $y=0$. The coefficients $C_{1}, C_{2},\cdots$ are all
determined from the equation of motion in terms of $C_0$ which can
have an arbitrary value. Here we see that $\delta\phi$ does not
vanish at the horizon, which makes $\phi_H$ different from
$\phi_0$ in general unlike the double-horizon extremal black hole.

From now on we consider only (nonnegative-)integer-power series
expansions of $\alpha^2$,\;$\beta$ and $\phi$. To solve the
equations of motion, we consider the following expansions and use
perturbative way. \bea
\alpha^2 &=& (1-\frac{r_{+}}{r})(1-\frac{r_{-}}{r})+(a_2 -1+\frac{3 r_{-}}{r_{+}})(\frac{r}{r_{+}}-1)^3+\cdots \nonumber \\
\cr\cr &=& (1-\frac{r_{-}}{r_{+}})(\frac{r}{r_+}-1)+(\frac{2
r_{-}}{r_{+}}-1)(\frac{r}{r_{+}}-1)^2
+a_{2}(\frac{r}{r_{+}}-1)^3+\cdots\\
\cr\cr \beta &=& r + r_{+}b_{2}(\frac{r}{r_{+}}-1)^2+
r_{+}b_{3}(\frac{r}{r_{+}}-1)^3+\cdots \nonumber \\
\cr\cr &=& r_+ +r_+(\frac{r}{r_{+}}-1) + r_{+}b_{2}(\frac{r}{r_{+}}-1)^2+ r_{+}b_{3}(\frac{r}{r_{+}}-1)^3+\cdots\\
\cr\cr \phi &=& \phi_{H} + \phi_{1}(\frac{r}{r_{+}}-1) +
\phi_{2}(\frac{r}{r_{+}}-1)^2 + \cdots\eea And we plug them into
the equations of motion and find the coefficients of the power
series expansions.

\begin{flushleft}
\underline{{\bf First order results}}
\end{flushleft}

\be (1-\frac{r_-}{r_+})\phi_1 = \frac{\gamma}{r_+^2}[Q_e^2 e^{2
\gamma \phi_H}-Q_m^2 e^{-2 \gamma \phi_H}] \ee We can express
reversely $\phi_{H}$ in terms $\phi_{1}$ and the charges, and find
easily that $\phi_{H}$ is a monotonically increasing function of
$\phi_{1}$. For larger $|\phi_1|$, $\phi_H$ deviates further from
$\phi_0$, as expected.

\begin{flushleft}
\underline{{\bf Second order results}}
\end{flushleft}

\bea a_2 &=& 1-\frac{3r_-}{r_+}+(1-\frac{r_-}{r_+})\phi_1^2\\
\cr\cr b_2 &=& -\frac{1}{2}\phi_1^2\\
\cr\cr 2(1-\frac{r_-}{r_+})\phi_2 &=&
-\phi_1+\frac{\gamma}{r_+^2}[(-1+\gamma\phi_1)Q_e^2 e^{2 \gamma
\phi_H}+(1+\gamma\phi_1)Q_m^2 e^{-2 \gamma \phi_H}]\eea

\begin{flushleft}
\underline{{\bf Higher order results}}
\end{flushleft}

Using $\alpha^2 \beta^2 = (r-r_+)(r-r_-)$ and the equations (45)
and (47), we can construct all the coefficients of $\phi-$series
in terms of $\phi_1$, where each coefficient $\phi_n (n\geq2)$ is
given by a polynomial of order $n$ of $\phi_1$ with $\phi_1$
undetermined by the equation of motion. Then from the equation
(46) all the coefficients of $\beta-$series can be written in
terms of $\phi_1$, where each coefficient $b_n (n\geq2)$ is given
by a polynomial of order $n$ of $\phi_1$. All the coefficients of
$\alpha^2-$series can appear in terms of $\phi_1$ from the
equation (44), where each coefficient $a_n (n\geq2)$ is a
polynomial of order $n$ of $\phi_1$. Finally the equation (43) can
be used for consistency check. Even if we change the basis of the
power series expansions from $(\frac{r}{r_+}-1)$ to
$(1-\frac{r_+}{r})$ where
$\frac{r}{r_+}-1=\sum(1-\frac{r_+}{r})^n$ all the coefficients
would be given as polynomials of $\phi_1$. So if we take the limit
$r\rightarrow\infty$, the asymptotic value $\phi_{\infty}$ is
determined by $\phi_1$, and vice versa\footnote{After change of
basis, $\phi_\infty$ can be shown to be \be \phi_\infty = \phi_H +
\sum^\infty_{n=1}\sum^n_{m=1}\frac{(n-1)!}{(m-1)!(n-m)!}\phi_m\ee.}.
Since $\phi_H$ is given in terms of $\phi_1$ and the fixed charges
of the black hole, it is determined by the asymptotic value
$\phi_{\infty}$ and the charges through the above processes. Thus
we observe unattractor phenomenon following this mechanism.

\section{Non-Supersymmetric Unattractors in Einstein-Born-Infeld Theories}

In this section, the Lagrangian, the metric ansatz and the
equations of motion are the same to those of section 3, and the
series expansions of $\alpha^2$,\;$\beta$ and $\phi$ have the same
form to those of section 4, but here the constant $\alpha_H$ is
defined by as follows: \be \alpha_H^2=1-2b r_+^2
e^{2\gamma\phi_H}\left(\sqrt{1+\frac{Q_e^2+Q_m^2e^{-4\gamma\phi_H}}{b
r_+^4}} -2 +
\frac{1}{\sqrt{1+\frac{Q_e^2+Q_m^2e^{-4\gamma\phi_H}}{b
r_+^4}}}\right)\ee And we can get the following result from the
Hamiltonian constraint at the zeroth order: \be
\alpha_H^2(1-\frac{r_-}{r_+})=1-2b r_+^2
e^{2\gamma\phi_H}\left(\sqrt{1+\frac{Q_e^2+Q_m^2e^{-4\gamma\phi_H}}{b
r_+^4}} -1 \right)\ee Dividing these two expressions,
 we can compute $(1-\frac{r_-}{r_+})$, and subtracting them shows that $(1-\frac{r_-}{r_+})$ is indeed less than
 1 for non-extremal black holes.

\begin{flushleft}
\underline{{\bf First order results}}
\end{flushleft}

\be  (1-\frac{r_-}{r_+})\phi_1 = \frac{2 b \gamma r_{+}^2 e^{2
\gamma \phi_H}}{\alpha_{H}^2}\left[-1+\frac{1+\frac{Q_e^2}{b
r_{+}^4}}{\sqrt{1+\frac{Q_e^2 + Q_m^2 e^{-4 \gamma \phi_H}}{b
r_{+}^4}}}\right] \ee Reversely $\phi_H$ can be presented by
$\phi_1$. With the same $V_{eff}$ as that of section 3, after
taking derivative with respect to $\phi$, we find $\phi_0$ is
given by \be e^{4\gamma\phi_0}=\frac{Q_m^2}{Q_e^2(1+\frac{Q_e^2}{b
r_{+}^4})}\ee After a little tedious calculus, $\phi_1$ can be
found to monotonically increase as $\phi_H$ increases only when
$\phi_H$ satisfies the following relation: \bea
B^2x^3+2BEx^2+2BEx+E^2 \nonumber \\  \cr
\geq\sqrt{Bx^2+E}(Bx^2+2BEx+E)\eea with \bea
B=4bQ_m^2\;\;\;;\;\;\;E=1+\frac{Q_e^2}{br_+^4}\;\;\;;\;\;\;x^{-1}=2br_+^2e^{2\gamma\phi_H}\;\;\;.
\eea However, in the limit $b\rightarrow\infty$ where the theory
approaches Einstein-Maxwell case, the above inequality is
satisfied automatically, and such bound does not exist as in
section 4, where $\phi_H$, the value of the scalar at the horizon,
monotonically increase according to $\phi_1$, the first
coefficient in $\phi$-series expansion, as consistent with our
intuition.

\begin{flushleft}
\underline{{\bf Second order results}}
\end{flushleft}

\bea a_2 &=& \frac{-4b r_{+}^2 e^{2 \gamma
\phi_H}}{3\alpha_H^2\sqrt{1+\frac{Q_e^2 + Q_m^2 e^{-4 \gamma
\phi_H}}{b r_+^4}}}[(1+\gamma\phi_{1})\left(\sqrt{1+\frac{Q_e^2 +
Q_m^2 e^{-4
\gamma \phi_H}}{b r_{+}^4}}-1\right)^2\nonumber \\
\cr\cr &&-\frac{Q_e^2 + (1+\gamma\phi_1)Q_m^2 e^{-4 \gamma
\phi_H}}{br_{+}^4\left(1+\frac{br_{+}^4}{Q_e^2 + Q_m^2 e^{-4
\gamma \phi_H}}\right)}]+1-\frac{3r_-}{r_+}+(1-\frac{r_-}{r_+})\phi_1^2 \\
\cr\cr b_2 &=& -\frac{1}{2}\phi_1^2\\
\cr\cr 2(1-\frac{r_-}{r_+})\phi_2 &=& -\phi_{1}+\frac{2 b \gamma
r_{+}^2 e^{2 \gamma \phi_H}}{\alpha_{H}^2}[ (1+\gamma
\phi_{1})\left(-1+\frac{1+\frac{Q_e^2}{b
r_{+}^4}}{\sqrt{1+\frac{Q_e^2 + Q_m^2 e^{-4 \gamma \phi_H}}{b
r_{+}^4}}}\right) \nonumber \\
\cr\cr &&+ \frac{\frac{Q_m^2 e^{-4 \gamma \phi_H}}{b
r_{+}^4}[1+\gamma\phi_{1}(1+\frac{Q_e^2}{b r_{+}^4})]}{\left(
1+\frac{Q_e^2 + Q_m^2 e^{-4 \gamma \phi_H}}{b r_{+}^4}
\right)^{\frac{3}{2}}}-\frac{\frac{Q_e^2}{b
r_{+}^4}}{\sqrt{1+\frac{Q_e^2 + Q_m^2 e^{-4 \gamma \phi_H}}{b
r_+^4}}}]\eea We can check that these coefficients reduce to the
values of Einstein-Maxwell theories in the limit
$b\rightarrow\infty.$

\begin{flushleft}
\underline{{\bf Higher order results}}
\end{flushleft}

From the equation (17), the coefficients $b_n (n\geq2)$ of
$\beta-$series are given in terms of $\phi_n (n\geq1)$ by simple
calculation. If we plug the power series into the rest three
equations, arbitrary $\phi_1$ determines all the other
coefficients. Actually, one equation out of the three is redundant
and can be used as consistency check of the results. So, in
principle, the asymptotic value of the scalar field $\phi_\infty$
can be expressed only by $\phi_1$ and the specific charges carried
by the Born-Infeld black holes, which can be evaluated after our
changing basis from $(\frac{r}{r_+}-1)$ to $(1-\frac{r_+}{r})$ and
taking the limit $r\rightarrow\infty$, and vice versa. Again the
unattractor behavior is identified.

\section{Conclusions}

In this paper, we studied unattractor behavior in a theory of
gravity coupled to a gauge field and a scalar field, with
Born-Infeld correction in the action. By investigating solutions
of the equations of motion, we observed the unattractor. We looked
for possible solutions which are regular
at the horizon and dependent on the asymptotic value of the scalar
field. The analysis of section 4 and 5 shows the behavior for
unattractor for non-extremal black holes in four-dimensional
asymptotically flat spacetime. Especially, we found the fact that
$\phi_H$, the value of the scalar at the horizon, is a
monotonically increasing function of $\phi_1$, the first
coefficient in $\phi$-series expansion only when $\phi_H$ has values in certain ranges which can be determined by the
derived inequality. It would be amusing to check whether this is true in black holes with other higher derivative correction terms, and whether $\phi_{\infty}$, the value of the scalar at the infinity, is a
monotonically increasing function of $\phi_1$.

\vskip 0.2cm

We used a perturbative approach to study the corrections to the
scalar field and take these backreaction corrections into the
metric, to show that the value of the scalar field at the infinity
indeed depends on the value of the scalar at the horizon, or
equivalently on $\phi_1$. However, unlike the case of a
Reissner-Nordstrom black hole, at higher orders in perturbative
theory, the metric components get backreaction corrections in the
Born-Infeld case. Fortunately, these corrections are small and
vanish at the horizon. We showed at asymptotic infinity, there are
different black hole solutions characterized by different values
taken by the scalar field of the theory. Near the horizon the
scalar field goes to specific values determined by its asymptotic
value and the charges of the black hole. It would be interesting
to generalize this analysis to the cases of asymptotic $AdS$ and
higher dimensional black holes.

\begin{center}
{\large {\bf Acknowledgements}}
\end{center}
The author would like to thank S.-J. Rey and H. Yavartanoo for
useful discussions and comments. This work was supported by the
Korea Research Foundation Leading Scientist Grant
(R02-2004-000-10150-0) and Star Faculty Grant
(KRF-2005-084-C00003).


\end{document}